              [1999/12/01 v1.4c Il Nuovo Cimento]
\documentclass{cimento}

 3
 2
 1
 1
 2

\def\med#1{<{#1}>_{2\p}}
\def\btab{\begin{tabular}}
\def\etab{\end{tabular}}
\def\grm{gravitomagnetic}

\def\Rfr#1{Eq.(\ref{#1})}
\def\Rfrs#1#2{Eqs.(\ref{#1})-(\ref{#2})}
\def\rfrs#1#2{eqs.(\ref{#1})-(\ref{#2})}
\def\l1{LAGEOS}
\def\l2{LAGEOS II}
\def\lt{Lense-Thirring}
\def\ra{{\mtc{R}}}

\def\la{{\mtc{L}}}
\def\rfr#1{eq.(\ref{#1})}
\def\dfa{\derp{\mtc{R}}{a}}
\def\dfm{\derp{\mtc{R}}{\mtc{M}}}
\def\dfog{\derp{\mtc{R}}{\og}}
\def\dfi{\derp{\mtc{R}}{i}}
\def\dfe{\derp{\mtc{R}}{e}}
\def\dfo{\derp{\mtc{R}}{\O}}
\def\cu{\rp{2}{na}}
\def\cd{\rp{1-e^2}{na^{2} e}}
\def\ctr{\rp{(1-e^2)^{1/2}}{na^{2} e}}
\def\cq{\rp{1}{na^{2}(1-e^2)^{1/2}\si}}

\def\bb#1{\bibitem{#1}}
\def\bar{\begin{array}}
\def\ear{\end{array}}
\def\eqi{\begin{equation}}
\def\eqf{\end{equation}}

\def\mtc#1{\mathcal{#1}}
\def\a{\alpha}

\def\z{\zeta}
\def\et{\eta}
\def\th{\theta}

\def\l{\lambda}
\def\m{\mu}
\def\n{\nu}
\def\x{\xi}

\def\p{\pi}
\def\r{\rho}

\def\f{\phi}

\def\og{\omega}

\def\D{\Delta}

\def\O{\Omega}
\def\co{\cos{\O}}
\def\so{\sin{\O}}
\def\cog{\cos{\og}}
\def\sog{\sin{\og}}
\def\ci{\cos{i}}
\def\si{\sin{i}}

\def\vass#1{\left\vert\ #1 \right\vert}
\def\rp#1#2{{#1\over#2}}

\def\km{\left(\matrix{\co\cog-\so\ci\sog & -\co\sog-\so\ci\cog & \so\si \cr
 \so\cog+\co\ci\sog & -\so\sog+\co\ci\cog & -\co\si
 \cr \si\sog & \si\cog & \ci \cr}\right)}

\def\derp#1#2{\rp{\partial{#1}}{\partial{#2}}}
\def\dert#1#2{\rp{{\rm d}{#1}}{{\rm d}{#2}}}

\def\ct#1{\cite{#1}}
\def\lb#1{\label{#1}}

\title{An alternative derivation of the Lense-Thirring\\
drag on the orbit of a test body}
\author{Lorenzo~Iorio\from{ins:x}}
\instlist{\inst{ins:x} Dipartimento Interateneo di Fisica dell' Universit{\`{a}} di Bari, Via Amendola
173, 70126, Bari}
\PACSes{\PACSit{95.30}{Relativity and gravitation}
        \PACSit{04.20}{Fundamental problems and general formalism}}
\begin{document}

\maketitle

\begin{abstract}
In the weak field and slow motion approximation of general
relativity a new approach in deriving the secular Lense-Thirring
effect on the orbital elements of a test body in the external
field of different central rotating sources exhibiting axial
symmetry is presented. The approach adopted in the present work,
in the case of a perfectly spherical source, leads to the well
known Lense-Thirring formulas for all the Keplerian orbital
elements of the freely falling particle. The corrections induced
in the case of a central nonspherical, axisymmetric body are also
worked out.
\end{abstract}

\section{Introduction.}
Latest years have seen increasing efforts devoted to the
measurement of the general relativistic  Lense-Thirring effect
\ct{ref:leti, ref:ciufw} in the weak gravitational field of the
Earth by means of artificial satellites. At present, there are two
main proposals which point towards the implementation of this
goal: the Gravity Probe-B mission \ct{ref:gpb}, and the approach
proposed in \ct{ref:ciuf1} which consists in using the actually
orbiting LAGEOS laser-ranged satellite and launching another
satellite of LAGEOS type, the LARES, with the same orbital
parameters of LAGEOS except for the inclination which should be
supplementary with respect to it. At present, both these two
satellites have not yet been launched: however, while the GP-B is
scheduled to fly in 2002, to date the LARES mission has not been
approved by any Spatial Agency.

Recently Ciufolini \ct{ref:ciuf2} has put forward an alternative
strategy based on the utilization of the already existing LAGEOS
and LAGEOS II  which could allow the detection of the
Lense-Thirring drag at a precision level of $20\ \%$
\ct{ref:ciuf2b, ref:ciuf3, ref:ciuf4}. While the GP-B mission  is
focused on the gravitomagnetic dragging of the spin of a freely
falling body, in the LAGEOS experiment it is the whole orbit of
the  satellite which is considered to undergo the secular
Lense-Thirring precession. More exactly, among the Keplerian
orbital elements the node and the perigee are affected by the
gravitomagnetic force. For LAGEOS and LAGEOS II its effect
consists in secular precessions amounting to:\begin{eqnarray} \dot
\O_{{\rm LT}}^{\rm LAGEOS }&\simeq& 31 \ {\rm mas/y},\lb{nodin}\\
\dot
\O_{{\rm LT}}^{{\rm LAGEOS II}}&\simeq& 31.5 \ {\rm mas/y}, \\
\dot \og_{{\rm LT}}^{{\rm LAGEOS II}}&\simeq& -57\ {\rm
mas/y},\lb{nodon}\end{eqnarray} where {\rm mas/y} stands for
milliarcseconds per year\footnote{The perigee of the LAGEOS has
not been considered since its rate is hard to measure
\ct{ref:ciuf2}.}

In this work it is presented an alternative strategy in order to
derive the secular gravitomagnetic precessions which reveals
itself useful especially in the prediction of the behavior of all
the Keplerian orbital elements of the test body in the
gravitational field of different kinds of central rotating
sources. Indeed, the calculations involve not only the secular
effects for a perfectly spherical source, which lead to
\rfrs{nodin}{nodon}, but also for a central body which is only
axially symmetric.

Until now the general relativistic motion in the field of a non spherical central body  has
been treated in \ct{ref:que1, ref:que2} for a static, nonrotating
source and in \ct{ref:teys1, ref:teys2, ref:adler} for a rotating source in the context of  the GP-B
mission. The influence of the non sphericity of the central source on the gravitomagnetic clock effect \ct{ref:mash1}
is investigated in \ct{ref:mash2}.

Our calculations are based on the
use of
the Lagrangian planetary equations
\ct{ref:vinti} and a
noncentral hamiltonian term
whose existence, if from
one hand can be rigorously deduced, from the other hand can be intuitively
guessed by
analogy from the corresponding term in
electrodynamics for a charged particle acted upon by  electric and magnetic fields.

\section{The gravitomagnetic potential}
In general, it can be proved that the general relativistic
equations of motion of a test particle of
 mass $m$ freely falling in a stationary
gravitational field, in the weak field and slow motion
approximation, are  given by \ct{ref:mash2}: \eqi m\rp{{\rm d}^2
{\bf r}}{{\rm d}t^2}\simeq m({\bf E}_{g}+\rp{1}{c}{\bf v} \times
{\bf B}_{g}). \label{cif}\eqf In \rfr{cif} $c$ is the speed of
light $in\ vacuo$ while ${\bf E}_{g}$ and ${\bf B}_{g}$ are the
gravitoelectric and the gravitomagnetic fields, respectively.

If a perfectly spherically symmetric rotating body is assumed as
gravitational source, in eq.(\ref{cif}) ${\bf E}_{g}=-G{M}{\bf
r}/{r^3}$ is the Newtonian gravitational
 field of a spherical body, with $M$ its mass and $G$ the Newtonian
gravitational constant,  while ${\bf B}_{g}$ is given by: \eqi
{\bf B}_{g}= \nabla \times {\bf A}_{g}\simeq 2\rp{G}{c}\ [\rp{
{\bf J}-3({\bf J}\cdot \hat{{\bf r}})\hat{{\bf
r}}}{r^3}]\lb{acca}, \eqf in which: \eqi {\bf A}_{g}({\bf
r})\simeq -2\rp{G}{c}\rp{{\bf J} \times{\bf
r}}{r^3}.\label{acchina}\eqf In \rfrs{acca}{acchina} {\bf J} is
the angular momentum of the central body. The field ${\bf A}_{g}$,
named gravitomagnetic potential, is due to the off-diagonal
components of the spacetime metric tensor: \eqi
g_{\m\n}=\et_{\m\n}+h_{\m\n},\ \m,\n=0,1,2,3,\eqf where
$\et_{\m\n}$ is the Minkowski metric tensor and $({\bf
A}_g)_k\equiv h_{0k},\ k=1,2,3$. In obtaining \rfr{acchina} a non
rotating
 reference frame $K\{x,\ y,\ z\}$
with the $z$ axis directed along  {\bf J} and the $\{x,\ y\}$
plane coinciding with the  equatorial plane  of the gravitational source has been used. The origin is
located at the center of mass of the central body.

The gravitomagnetic potential
generates a non central
gravitational contribute due uniquely to the angular momentum of the gravitational
source
 that the Newtonian mechanics does not predict, though the conditions of
validity of  eq.(\ref{cif}) are the same for which the latter
holds as well.
%Eq.(\ref{e}) and (\ref{h}) are also valid
%if the test particle is far away from a non spherical rotating body \ct{leti}, because if
%the distance is great enough all the effects of the non sphericity in the
%gravitational field vanish.
%On the other hand, in the electromagnetism, if a localized stationary current distribution is
%considered, far from it the potential vector is given, at first
%order, by
%\ct{jack}:
%\eqi {\bf A}({\bf r})\simeq \rp{{{\bf m}}\times {\bf r}}{r^3} \lb{a}\eqf
%where ${{\bf m}}$ is the magnetic moment of the current distribution.
%Consequently, \eqi {\bf B}=\nabla \times {\bf A}\simeq
%\rp{ 3({{\bf m}}\cdot\hat{{\bf r}})\hat{{\bf r}}-{{\bf m}}}{r^3}\lb{b}\eqf
%Comparing eq.(\ref{h}) and eq.(\ref{gm}) with eq.(\ref{a}) and eq.(\ref{b}),
%it appears that the angular momentum ${\bf J}$ of a spinning massive object
%plays the same role of the magnetic moment ${{\bf m}}$
%of a stationary current distribution.
So it is possible to speak of
mass-energy currents whose motion exerts a non central
 gravitational force on a test
massive body analogous to the  Lorentz force felt by a charged
particle of charge $q$ and mass $m$ when it is moving in a
electromagnetic field. Indeed, its equations of motions: \eqi
m\rp{{\rm d}^2 {\bf r}}{{\rm d}t^2}\simeq q({\bf E}+\rp{1}{c}{\bf
v} \times {\bf B}) \label{lor}\eqf are formally analogous to
eq.(\ref{cif}). Eq.(\ref{lor}) can be derived  by means of the
Lagrangian: \eqi {\mtc{L}}_{e.m.}=\rp{1}{2}mv^2-qV+ \rp{q}{c}({\bf
v}\cdot{\bf A}),\lb{ele}\eqf where ${\bf v}$ is the velocity of
the particle while $V$ and {\bf A} are the  scalar and vector
potential, respectively, of the electromagnetic field.

Since one of the most promising way to detect the gravitomagnetic precession
consists in employing artificial Earth satellites, it would be helpful to
derive the rate equations for the change in the parameters that characterize
their orbits. To this aim one could
 introduce ``by hand'' a perturbative
 term $k\ ({\bf v}\cdot{\bf A}_{g})$ in the gravitational Lagrangian of the particle and use it in
some particular perturbative scheme;
 the  constant $k$ would be determined by means
of dimensional considerations and taking in account that it should
be built up of universal constants. In fact it is possible to show
that a non central term analogous to $\rp{q}{c}({\bf v}\cdot{\bf
A})$ can be rigorously deduced in the Lagrangian of  a test body
in the gravitational field of a spinning mass-energy distribution,
and that it can be exploited in deriving straightforwardly the
effect of the gravitomagnetic potential on the Keplerian orbital
elements of the test body.\section{The rate equations for the
Keplerian orbital elements} The relativistic Lagrangian for a free
particle of mass $m$ in a gravitational field  can be cast into
the form:
\eqi{\mtc{L}}={\mtc{L}}^{(0)}+{\mtc{L}}^{(1)}\lb{lag2}.\eqf In
\rfr{lag2} the term ${\mtc{L}}^{(1)}$ refers to the off-diagonal
terms of the metric: \eqi {\mtc{L}}^{(1)}
=\rp{m}{c}g_{0k}\dot{x^{0}}\dot{x^{k}}.\lb{perini}\eqf In this
case, recalling that the slow motion approximation is used,
\rfr{perini} becomes \ct{ref:mash2}: \eqi
{\mtc{L}}^{(1)}\equiv{\mtc{L}}_{gm}\simeq \rp{m}{c}\ ({\bf
A}_{g}\cdot{\bf v}).\lb{cime}\eqf In this paper it is proposed to
adopt ${\mtc{L}}_{gm}$ given by \rfr{cime}, with ${\bf A}_{g}$
given by \rfr{acchina}, in order to derive the Lense-Thirring
effect on the orbital elements of a particle in the field of a
rotating gravitational source.

To this aim it must be assumed that under the gravitomagnetic force the departures
of the test body' s trajectory from the unperturbed Keplerian ellipse
are very small
in time. This allows to introduce the concept of osculating ellipse.
So the perturbed motion can be described in terms
of unperturbed Keplerian elements varying in time.
Consider the frame $K\{x,\ y,\ z\}$ previously defined and a frame
 $K^{'}\{X,
\ Y,\ Z\}$ with the $Z$ axis directed along the orbital angular
momentum ${\bf l}$ of the test body, the plane $\{X, \ Y\}$
coinciding with the orbital plane of the test particle and the $X$
axis directed toward the pericenter. $K\{x,\ y,\ z\}$ and $K^{'}
\{X, \ Y,\ Z\}$  have the same origin located in the center of
mass of the central body.
The Keplerian orbital elements are:\\
$\bullet\ \  a,\ e$\\
The semimajor axis and the ellipticity which define the size and the shape of
the orbit in its plane.\\
$\bullet\ \  \O,\ i$\\
The longitude of the ascending node and the inclination which fix
the orientation of the orbit in the space, i.e. of $K^{'}\{X, \
Y,\ Z\}$ with respect to $K\{x,\ y,\ z\}$. The
 longitude of the ascending node $\O$ is the angle in the equatorial plane of
$K\{x,\ y,\ z\}$ between the $x$ axis and the line of nodes in which the orbital
plane intersects the equatorial plane. The inclination $i$ is the angle between
$z$ and $Z$ axis.\\
$\bullet\ \  \og,\ \mtc{M}$\\
The argument of pericenter and the mean anomaly. The argument of
pericenter $\og$ is the angle in the orbital plane between the
line of nodes and the $X$ axis; it defines the orientation of the
orbit in its plane. The mean anomaly $\mtc{M}$ specifies the
motion of the test particle on its orbit. It is related to the
mean motion $n=(GM)^{1/2}a^{-3/2}$ through ${\cal{M}}=n(t-t_{p})$
in which $t_{p}$ is the time of pericenter passage.\\It is
customary to define also the longitude of pericenter\\ $\bullet\
 \varpi=\O+\og$,\\
the argument of latitude\\ $\bullet\ u=\og+f$\\ where $f$ is the angle, called true anomaly,
which in the orbital plane determines the position of the test particle with
respect to the pericenter, and the mean longitude at the
epoch $t_0$ \\ $\bullet\
 \varepsilon=\varpi+n(t_0-t_p)$. If $t_0=0$, it is customary to write
$\varepsilon$ as
$L_0=\varpi-n t_p$.
\\ The matrix ${\bf R}_{xX}$ for
 the change of coordinates from
$K^{'} \{X, \ Y,\ Z\}$ to $K\{x,\ y,\ z\}$ is: \eqi \km
\lb{mat}.\eqf Using eq.(\ref{mat}) and $X=r\cos{f}$, \ \
$Y=r\sin{f}$, \ \ $Z=0$ it is possible to express the geocentric
rectangular Cartesian coordinates of the orbiter in terms of its
Keplerian elements:\eqi\left\{\begin{array}{l}x=r (\cos{u}\co-
\sin{u}\ci\so)\\
y=r(\cos{u}\so+
\sin{u}\ci\co)\\
z=r\sin{u}\si.   \lb{z} \end{array}\right.\eqf
Redefining suitably the origin of the angle $\O$ so that $\co=1$, $\so=0$, the
previous equations become:
\eqi\left\{\begin{array}{l}x=r\cos{u}\\y=r\sin{u}\ci\\z=
r\sin{u}\si.\lb{paolozzi}\end{array}\right.\eqf
Considering for the test particle  the total mechanical energy with the sign
reversed, according to \ct{ref:vinti}, ${\mtc{F}} \equiv
 -{\mathcal{E}}_{tot}=-(\mathcal{T}+
\mathcal{U})$, where $\mathcal{T}$ and $\mathcal{U}$ are the kinetic and
potential energies per unit mass, it is possible to work out the analytical
expressions for the rate of changes of $a,\ e,\ i,\ \O,\ \og,\ {\mtc{M}}$ due
to any non central gravitational contribution. To this aim
it is useful isolating in $\mathcal{U}$ the central part $\mathcal{-C}$ of the gravitational
field from the terms $\mathcal{-R}$ which
may cause the Keplerian orbital elements to change in time:
${\mathcal{U}}=-{\mtc{C}}-{\mathcal{R}}$.  In this way $\mtc{F}$ becomes:
\eqi {\mtc{F}}=\rp{GM}{r}+{\mathcal{R}}-{\mathcal{T}}=\rp{GM}{2a}+
{\mathcal{R}}.\lb{kau}\eqf
Concerning the perturbative scheme to be employed,
the well known Lagrange planetary
equations  are adopted.
%The starting point is the set of the following
%equations:
%\begin{eqnarray}\dert{x_k}{t}&=&\dot{x_k},\ k=1,2,3,\\
%\dert{\dot{x_k}}{t}&=&-\derp{{\mtc{U}}}{x_k},\ k=1,2,3,\end{eqnarray}
%where ${x_k},\ k=1,2,3$ are the geocentric rectangular Cartesian coordinates of
%the test body. Considering them as functions of the six orbital elements,
%denoted by $s_r,\ r=1,..,6$, it is possible to obtain:
%\eqi \sum_{r=1}^{6}[s_{h},\ s_{r}]\dert{s_{r}}{t}=\derp{{\mtc{F}}}{s_{h}},\
%h=1,..,6\eqf where the Lagrange brackets:
%\eqi[s_{h},\ s_{r}]=\sum_{i=1}^{3}(\derp{x_i}{s_{h}}\derp{\dot{x_i}}{s_{r}}
%-\derp{\dot{x_i}}{s_{h}}\derp{x_i}{s_{r}}), h,r=1,..,6\eqf
%are used.
At first order, they are:
\begin{eqnarray}\dert{a}{t}&=& \cu \ \dfm ,\lb{dos}\\
\dert{e}{t}&=& \cd \ \dfm - \ctr \ \dfog ,\lb{ee}\\
\dert{i}{t}&=& \ci \cq \ \dfog - \cq \ \dfo ,\\
\dert{\O}{t}&=& \cq \ \dfi ,\lb{omeg}\\
\dert{\og}{t}&=&- \ci \cq \ \dfi + \ctr \ \dfe ,\lb{omeghin}\\
\dert{\mtc{M}}{t}&=&n- \cd \ \dfe - \cu \ \dfa
.\lb{tres}\end{eqnarray} The idea of this work consists in using
${\mtc{L}}_{gm}$ to obtain a suitable non central term
${\mtc{R}}_{gm}$ to be employed in these equations. This can be
done considering the Hamiltonian for the test particle: \eqi
{\mtc{H}}={\bf p}\cdot{\bf v}-{\mtc{L}}.\lb{ham}\eqf Inserting
\rfr{lag2} in \rfr{ham} one has: \eqi
{\mtc{H}}={\mtc{H}}^{(0)}+{\mtc{H}}_{gm},\eqf with
${\mtc{H}}_{gm}=-\rp{m}{c}\ ({\bf A}_{g}\cdot{\bf v})$. So it can
be posed:\eqi {\mtc{R}}_{gm}=-\rp{{\mtc{H}}_{gm}}{m}=\rp{1}{c}\
({\bf A}_{g}\cdot{\bf v})\lb{r}.\eqf

Now it is useful to  express \rfr{r} in terms of the Keplerian
elements. Referring to eq.(\ref{acchina}), \rfr{paolozzi},
recalling that in the frame $K\{x,\ y,\ z\}$ ${\bf J}=(0,\ 0,\ J)$
and that for an unperturbed Keplerian motion: \eqi
\rp{1}{r}=\rp{(1+e\cos{f})}{a(1-e^2)}, \eqf it is possible to
obtain, for a perfectly spherical central body: \eqi
{\mtc{R}}_{gm}= -\rp{2G}{c^2}\rp{J\ci}{r} \dot{u}=
-\rp{2GJ\ci}{c^2}\rp{(1+e\cos{f})}{a(1-e^2)}\dot{u}.\lb{ru}\eqf In
\rfr{ru}  $\dot u\simeq \dot f$ is assumed due to
 the the fact that the osculating
element $\og$ may be retained almost constant on the temporal scale of variation of the
true anomaly of the test body.\section{Secular gravitomagnetic
effects on the Keplerian orbital elements: spherical central source}

The secular effects can be worked out by adopting the same
strategy followed in \ct{ref:vinti} for a similar kind of
perturbing functions. When eq.(\ref{ru}) is averaged over one
orbital period $P$ of the test body, $a,\ e ,\ i,\ \O$ and $\og$
are to be considered constant:
\eqi\med{{\mtc{R}}}=-\rp{1}{P}\int_{0}^{P}
\rp{G}{c^2}\rp{2J}{r}\ci du
 =-2n\rp{G}{c^2}\rp{J\ci}{ 2\p}\int_{0}^{2\p}
\rp{df}{r}=-2n\rp{G}{c^2}\rp{J\cos{i}}{
a(1-e^2)}\lb{fine}.\label{po}\eqf The relation $du=d\og+df=df$ has
been used in \rfr{po} which can now be used in determining the
secular changes of the Keplerian orbital elements of the test
body. It yields:
\begin{eqnarray}
\derp{\med{\mtc{R}}}{{\mtc{M}}}&=&0,\lb{soluz}\\
\derp{\med{\mtc{R}}}{\og}&=&0,\\
\derp{\med{\mtc{R}}}{\O}&=&0,\\
\derp{\med{\mtc{R}}}{i}&=&
2n\rp{G}{c^2}\rp{J\sin{i}}{a(1-e^2)},\\
\derp{\med{\mtc{R}}}{e}&=&-4n
\rp{G}{c^2}\rp{Je\cos{i}}{a(1-e^2)^2}\lb{re}.
\end{eqnarray}
A particular care is needed for the treatment of $n$ when the
derivative of $\med{\mtc{R}}$ with respect to $a$ is taken;
indeed, it must be posed as:
\eqi\derp{\med{\mtc{R}}}{a}=\derp{\med{\mtc{R}}}{a}_{\vert_{n}}+
\derp{\med{\mtc{R}}}{n}_{\vert_{a}}\derp{n}{a}.\lb{palle}\eqf In
\rfr{palle} \eqi\derp{\med{\mtc{R}}}{a}_{\vert_{n}}= 2
n\rp{G}{c^2}\rp{J\cos{i}}{a^2(1-e^2)},\lb{uffa}\eqf and
\eqi\derp{\med{\mtc{R}}}{n}_{\vert_{a}}\derp{n}{a}= 3
n\rp{G}{c^2}\rp{J\cos{i}}{a^2(1-e^2)}.\eqf From \rfr{dos} and
\rfr{soluz} it appears that there are no secular changes in the
semimajor axis, and so the orbital period of the test body,
related to the mean motion by $P=2\p/n$, can be considered
constant. This implies that in \rfr{palle} only \rfr{uffa} must be
retained. Using  \rfrs{soluz}{palle} in  \rfrs{dos}{tres} one
obtains for the secular rates:
\begin{eqnarray}
\left.\dert{{a}}{t}\right|_{{\rm LT}}&=& 0,\lb{alt}\\
\left.\dert{{e}}{t}\right|_{{\rm LT}}&=& 0,\\
\left.\dert{{i}}{t}\right|_{{\rm LT}}&=& 0,\\
\left.\dert{{\O}}{t}\right|_{{\rm LT}}&=&\rp{G}{c^2}\rp{2J}{a^{3}(1-e^2)^{3/2}},\\
\left.\dert{{\og}}{t}\right|_{{\rm LT}}&=&-\rp{G}{c^2}\rp{6J}{a^{3}(1-e^2)^{3/2}} \ci,\\
\left.\dert{{{\mtc{M}}}}{t}\right|_{{\rm
LT}}&=&0\lb{moo}.\end{eqnarray} They are the well known
Lense-Thirring equations \ct{ref:leti, ref:ciufw}. In deriving
them it has been assumed that the spatial average over $f$ yields
the same results for the temporal average \ct{ref:milani}.
%Note that if one takes the temporal average
%over an
%orbital period of \rfr{dat}-\rfr{miam}, this leads to
%\rfr{alt}-\rfr{moo} since
%\eqi \rp{1}{P}\int_{0}^{P}fdf=n,\
%\rp{1}{P}\int_{0}^{P}\sin{f}df=\rp{1}{P}\int_{0}^{P}\cos{f}df=0
%.\eqf
%5Our strategy allows to obtain \rfr{alt}-\rfr{moo} in a very straightforward manner and points out clearly that
% %for the mean
%anomaly there are no secular changes due to the Lense-Thirring effect (cfr.
%\ct{leti} in which the Gauss perturbative equations and the mean longitude at the epoch
% $t_0=0$ are used).
%\\
%
%The physical meaning of the previous equations is more clear if the whole of the
%orbital plane of the test particle is considered like an enormous gyroscope whose
%angular momentum ${\bf l}$ tends to preserve its orientation with respect to
%$K\{x,\ y,\ z\}$. This is what it would happen if only central forces were
%applied on it. But the gravitomagnetic force $\rp{m}{c}({\bf v})\times{\bf B}_{g})$ is not
%collinear with the position vector ${\bf r}$ of the particle and so it
%generates a momentum which force ${\bf l}$ to undergo a precession.
%Consequently, the orientation of the orbit and the elements which determine it
%change with time. The variation of $\og$ is due to the general
% fact that the Runge-Lenz
%vector lying in the orbital plane is not conserved when the force field is
%not central \ct{bern}.
\section{Secular
gravitomagnetic effects on the Keplerian orbital elements: nonspherical central source}
In this Section
we shall deal with a
non spherical central rotating source with axial symmetry around the rotation axis.

In classical electrodynamics the potential vector for a generic steady current
distribution can be written as:\eqi {\bf A({\bf r})}=\rp{1}{c}\int \rp{\r({\bf
r^{'}}){\bf v}}{\vass{{\bf r}-{\bf r}^{'}}}d{\bf r^{'}}.\lb{tre}\eqf
The quantity $\r$ is the charge density which, in general, depends
on ${\bf r}$, but, in this case, not on time.
If we assume that the current distribution rotates uniformly around an axis,
chosen as $z$ axis,
then, since for any
current element $
{\bf v}=\a \hat{z}\times{\bf r}$ with $\a$ angular velocity of the
current
distribution \rfr{tre} becomes:\eqi
{\bf A({\bf r})}=\rp{\a}{c}\int \rp{\r({\bf
r^{'}})\hat {z} \times{\bf r}}{\vass{{\bf r}-{\bf r}^{'}}}d{\bf
r^{'}}.\lb{ops}\eqf
In order to deal with an axisymmetric  current distribution let us
introduce the cylindrical coordinates:
\eqi\left\{\begin{array}{l}x=\x\cos{\f}\\
y=\x\sin{\f}\\
z=z,  \lb{cil} \end{array}\right.\eqf $\x \geq 0,\ 0\leq \f < 2\p$.
In this case it can be shown that the lines of the potential vector are
circles around the $z$ axis and its modulus is a cylindrical symmetric
function of $\x$ and $z$. So, \rfr{ops} reduces to:
\eqi {\bf A}({\bf r})=\rp{\a}{c}H(\x,\ z){\bf e}_{\f}.\lb{quater}\eqf The
function $H(\x,\ z)$ remains unchanged under rotations around the $z$ axis. Passing from
cylindrical  to rectangular Cartesian coordinates, the components of the
potential vector become:
\begin{eqnarray}
A_1 &=& -\rp{\a}{c}H(\x,\ z)y,\lb{cin}\\
A_2 &=& \rp{\a}{c}H(\x,\ z)x,\lb{sei}\\
A_3 &=& 0\lb{null}.
\end{eqnarray}  Recalling that, in the linear approximation, the general relativistic
{\grm} potential can be obtained from the vector potential of
electromagnetism times $-4G$ \ct{ref:ciufw}, \rfrs{cin}{null} lead
to:
\begin{eqnarray}
h_{01} &=& \rp{4G\a}{c}H(\x,\ z)y,\lb{sette}\\
h_{02} &=& -\rp{4G\a}{c}H(\x,\ z)x,\lb{otto}\\
h_{03} &=& 0\lb{undici}.
\end{eqnarray}
Of course, \rfrs{sette}{undici} can be rigorously obtained solving
the linearized Einstein field equation, written in the Lorenz
gauge, for a localized ($g_{\m\n} \rightarrow \et_{\m\n}$ at
spatial infinity), axisymmetric, stationary mass distribution in
the weak field and slow motion approximation:
\begin{eqnarray} \D h_{0k}&=&0,\  k=1,2,3\lb{nove}\\
\D h_{0k}&=&\rp{16\p G}{c^4}T_{0k},\ k=1,2,3\lb{dodici}.\end{eqnarray} The quantities
$T_{0k}$ are the $\{0k\}$ components of the stress-energy tensor for the matter;
in deriving \rfr{dodici} the internal stresses have been neglected. \Rfr{dodici} is valid
inside the matter, while \rfr{nove}
holds in the free space outside the central body and
 tells us that $H(\x,\ z)y$ and $H(\x,\ z)x$
 are harmonic functions. This feature was used by Teyssandier in
obtaining a multipolar expansion of $H(\x,\ z)$ \ct{ref:teys1}. Introducing in the
frame $K$ the usual
spherical coordinates $\{r,\ \th,\ \f\}$, if $r_e$ is the radius of the
smallest sphere centered on the origin of the coordinates containing the whole
body (in practice, it should be the equatorial radius, $R_{\oplus}$ in the case of the Earth), in
the region $r\geq r_1 > r_e$ $H(r,\ \th)$ is given by:
\eqi H(r,\ \th)=\rp{I}{2r^3}\ [ 1-\sum_{l=1}^{\infty} K_l (\rp{r_e}{r})^l
P_{l+1}^{'} (\cos{\th})],\lb{yy}\eqf with:\eqi K_l=\rp{2}{2l+3}\ \rp{M r_e^2}{I}(
L_l-J_{l+2}),\lb{k}\eqf
\eqi L_l=-\rp{1}{M r_e^{l+2}}\int \r(r^{'},\
\th^{'}) {r^{'}}^{l+2} P_l(\cos{\th^{'}})d{\bf
r}^{'}\lb{l}.\eqf In \rfrs{yy}{l} $I$ is the moment of inertia of the body about
the $z$ axis, $M$ is
the total mass of the
central body, $\r$ is
its density, $P_{l+1}^{'} (\cos{\th})$ is the first derivative of the Legendre
polynomial of degree $l+1$ and $J_l$ is the
 Newtonian multipole moment of degree $l$,
given by: \eqi J_l=-\rp{1}{M r_e^l}\int
\r(r^{'},\
\th^{'}) {r^{'}}^{l} P_l(\cos{\th^{'}})d{\bf
r}^{'}.\lb{j}\eqf It is interesting to note that, if in \rfr{yy} only the
spherical symmetric term:
\eqi H^{(0)}(r)=\rp{I}{2r^3}\eqf is retained, as it would be the case for
a perfectly spherical central body, \rfrs{sette}{undici} reduce to:
\begin{eqnarray}
h_{01}^{(0)} &=& \rp{2GI\a}{c}\rp{y}{r^3},\lb{h1}\\
h_{02}^{(0)} &=& -\rp{2GI\a}{c}\rp{x}{r^3},\lb{h2}\\
h_{03}^{(0)} &=& 0\lb{h3}.
\end{eqnarray}
For a spherical rotating body $J=I\a$, and so \rfrs{h1}{h3} can be
cast into the familiar form:\eqi {\bf A}_{g}^{(0)}=
-\rp{2G}{c}\rp{{\bf J}\times {\bf r}}{r^3}.\eqf

The correction of order $l$ to the {\grm} potential due to the nonsphericity of
the central rotating body is given by:
\begin{eqnarray}
h_{01}^{(l)} &=& -\rp{2GI\a}{r^3 c^3}\ y\ K_l (\rp{r_e}{r})^l
P_{l+1}^{'} (\cos{\th}),\lb{hl1}\\
h_{02}^{(l)} &=& \rp{2GI\a}{r^3 c^3}\ x\ K_l (\rp{r_e}{r})^l
P_{l+1}^{'} (\cos{\th}),\lb{hl2}\\
h_{03}^{(l)} &=& 0\lb{hl3}.
\end{eqnarray}
Modeling the central body as a spheroid stratified into ellipsoidal shells, in
\ct{ref:teys1} is shown that all the relativistic coefficients $K_l$ are null,
except $K_2$; if we consider our planet, from the analysis of Teyssandier
of various models of Earth's interior,
 it results to be positive and of ${\mtc{O}}(10^{-3})$.

The starting point in deriving the relativistic multipole
corrections of degree $l$ to \rfrs{alt}{moo} is the perturbative
lagrangian term: \eqi \la^{(l)}_{gm}=\rp{m}{c}({\bf
A}_{g}^{(l)}\cdot {\bf v}),\eqf where $m$ is the mass of the point
particle and ${\bf v}$ is its velocity. In the case $l=2$, in
\rfrs{hl1}{hl3} the first derivative of the Legendre polynomial of
degree
 $3$ appear;
\begin{eqnarray}
 P_3(q)&=&\rp{1}{2}\ (5{q}^3-3q),\ q=\cos{\th},\\
 P^{'}_3(q)&=&\rp{1}{2}\ (15{q}^2-3),\
q=\cos{\th}.\lb{cs}
\end{eqnarray}
 Inserting \rfr{cs} in \rfrs{hl1}{hl3} for $l=2$ leads to:
\eqi \la^{(2)}_{gm}=-m\rp{GI\a}{c^2 r^3}\ \rp{(y\dot x-x\dot
y)}{r^2}\ r^2_{e}K_2 (15{\cos^2{\th}}-3).\lb{l2}\eqf Such a
perturbative term must be expressed in term of the Keplerian
orbital elements in order to be employed in the first order
Lagrange planetary equations. In this case $\ra$ is
$\ra^{(2)}=\rp{1}{c}({\bf A}_{g}^{(2)}\cdot {\bf v})$. Using
\rfr{paolozzi} and recalling that $z=r\cos{\th}$ it is possible to
obtain:
 \begin{eqnarray}
 \rp{(y\dot x-x\dot y)}{r^2}&=& -\dot u \cos{i}
\lb{3}\\
\cos{\th}&=&\sin{u}\sin{i}.\lb{cos}
\end{eqnarray}
 Neglecting terms of order ${\mtc{O}}(e^n),\ n \geq 2$,
we can write:\eqi
\rp{1}{r^3}\simeq\rp{1+3e\cos{f}}{a^3(1-e^2)^3}\lb{4},\eqf having
assumed: \eqi \rp{1}{r}=\rp{1+e\cos{f}}{a(1-e^2)}\lb{new}.\eqf
\Rfrs{3}{4} in \rfr{l2} give: \eqi\ra^{(2)}=\rp{GI\a}{c^2}
\rp{r^2_{e}K_2\cos{i}}{a^3(1-e^2)^3}(1+3e\cos{f}) (15{\sin^2{u}} \
{\sin^2{i}} -3)\dot u.\lb{r2}\eqf If we want to investigate the
secular trends of the orbital elements, we must average \rfr{r2}
over an orbital period of the test body, as done in the previous
Section. It is straightforward to obtain: \eqi \med{(1+3e\cos{f})
(15{\sin^2{u}}\ {\sin^2{i}} -3)\ \dot
u}=\rp{n}{2}(9-15{\cos^2{i}})\lb{media}. \eqf In deriving
\rfr{media} we have adopted the reasonable assumption that the
pericenter of the test body remains almost unchanged during an
orbital revolution, i. e. $du=d\og+df\simeq df$. So we have:\eqi
\med{\ra^{(2)}}=n\rp{GI\a}{2c^2}
\rp{r^2_{e}K_2}{a^3(1-e^2)^3}(9\cos{i}-15{\cos^3{i}}).\lb{rmedia}\eqf
\Rfr{rmedia} can be considered as the 2nd order correction to the
gravitomagnetic perturbing function given in \rfr{po}.
\Rfr{rmedia} yields:
%%%%%%%%%%%%%%%%%%%%%%%%%%%%%%%%%
\begin{eqnarray}
\derp{\med{\ra^{(2)}}}{{\mtc{M}}}&=&0,\lb{rot1}\\
\derp{\med{\ra^{(2)}}}{\og}&=&0,\\
\derp{\med{\ra^{(2)}}}{\O}&=&0,\\
\derp{\med{\ra^{(2)}}}{i}&=&n\rp{GI\a}{2c^2}
\rp{r^2_{e}K_2}{a^3(1-e^2)^3}\sin{i}(45{\cos^2{i}}-9) ,\\
\derp{\med{\ra^{(2)}}}{e}&=& n\rp{3eGI\a}{c^2}
\rp{r^2_{e}K_2}{a^3(1-e^2)^4}(9\cos{i}-15{\cos^3{i}}),\\
\derp{\med{\ra^{(2)}}}{a}_{\vert_{n}}&=&-n\rp{3GI\a}{2c^2}
\rp{r^2_{e}K_2}{a^4(1-e^2)^3}(9\cos{i}-15{\cos^3{i}}).\lb{rot2}
\end{eqnarray}
%%%%%%%%%%%%%%%%%%%%%%%%%%%%%%%%%%
By inserting \rfrs{rot1}{rot2} in \rfrs{dos}{tres} it can be
obtained for the secular rates:
\begin{eqnarray}
\left.\dert{{a}}{t}\right|^{(2)}_{{\rm LT}}&=& 0,\lb{alto}\\
\left.\dert{{e}}{t}\right|^{(2)}_{{\rm LT}}&=& 0,\\
\left.\dert{{i}}{t}\right|^{(2)}_{{\rm LT}}&=& 0,\\
\left.\dert{{\O}}{t}\right|^{(2)}_{{\rm LT}}&=&\rp{GI\a}{2c^2}
\rp{r^2_{e}K_2}{a^5(1-e^2)^{7/2}}(45{\cos^2{i}}-9),\lb{olt2}\\
\left.\dert{{\og}}{t}\right|^{(2)}_{{\rm LT}}&=&
-\cos{i}\left.\dert{{\O}}{t}\right|^{(2)}_{{\rm LT}}+
\rp{3GI\a}{c^2}
\rp{r^2_{e}K_2}{a^5(1-e^2)^{7/2}}(9\cos{i}-15{\cos^3{i}}),\lb{oglt2}\\
\left.\dert{{{\mtc{M}}}}{t}\right|^{(2)}_{{\rm LT}}&=&0\lb{mlt}.
\end{eqnarray}
\Rfrs{alto}{mlt} yield the
corrections to the precessional
rate Lense-Thirring equations when the central rotating body is not perfectly spherical but only axially
symmetric.

If we use in \rfrs{olt2}{oglt2} the values $r_e=R_{\oplus}\simeq
6378$ km, $K_2=0.874\times 10^{-3}$ \ct{ref:teys2} we can obtain
an estimate of the sensitivity of LAGEOS  and LAGEOS II to the
relativistic Earth's quadrupole correction to the {\lt}
precessional rates. The results are:
\begin{eqnarray}
\dot \O^{(2)}_{{\rm LAGEOS}}&=&-6.7\times 10^{-1}\ {\rm mas/century},\\
\dot \O^{(2)}_{{\rm LAGEOS II}}&=&2.8\ {\rm mas/century},\\
\dot \og^{(2)}_{{\rm LAGEOS II}}&=&5.4\ {\rm mas/century}.
\end{eqnarray}
The present accuracy in the measurement of the LAGEOSs' rates of
the node and the perigee is of the order of 1 {\rm mas/y}. This
implies that such tiny corrections do not affect the current
efforts in detecting the Lense-Thirring drag.
\section{Conclusions}
In the weak field
and slow motion approximation
of general relativity we have developed an alternative approach to the calculation of
the Lense-Thirring effect
on the Keplerian orbital
elements of a test particle freely falling in the field of different kinds of
axially symmetric central sources.

Such strategy  stresses the
formal analogy  with the electromagnetism recalling the lagrangian of a
charged
particle
acted upon by the Lorentz force
and create a link to space geodesy by exploiting  the widely used Lagrangian planetary equations.

For a perfectly spherical
source the well known Lense-Thirring
precessional rates for the node and the perigee are obtained.

If
departures from sphericity
of the gravitational source are
accounted for, it has been found that only the node and the perigee are
affected by them through additional secular rates.

At present they cannot influence the current measurement of the Lense-Thirring drag by means of LAGEOS
and LAGEOS II because they fall below the  experimental sensitivity.

\acknowledgments
I am profoundly indebted to I. Ciufolini who focused my attention to the fascinating
topic of gravitomagnetism.
I would like also to thank M. Gasperini for discussions, P. Colangelo who has read
carefully the manuscript and L. Guerriero who supported me at Bari.

\end{document}